\documentclass[%
 reprint,
 amsmath,amssymb,
 aps, 
]{revtex4-1}

\usepackage{textcomp}
\usepackage{graphicx}
\usepackage{dcolumn}
\usepackage{bm}
\usepackage{varioref}

\usepackage{hyperref}

\usepackage{natbib}

\usepackage{color}
\usepackage{soul}

\usepackage{booktabs}

\usepackage{pifont}
\begin{document}

\title{Thermal transport in crystals as a kinetic theory of relaxons} 

\author{Andrea Cepellotti}
\author{Nicola Marzari}
\email{nicola.marzari@epfl.ch}
\affiliation{Theory and Simulations of Materials (THEOS), and National Centre for Computational Design and Discovery of Novel Materials (MARVEL), \'Ecole Polytechnique F\'ed\'erale de Lausanne, Station 9, 1015 Lausanne, Switzerland}

\begin{abstract} 
Thermal conductivity in dielectric crystals is the result of
the relaxation of lattice vibrations described by the phonon Boltzmann
transport equation. Remarkably, an exact microscopic definition of the heat
carriers and their relaxation times is still missing: phonons, typically
regarded as the relevant excitations for thermal transport, cannot be
identified as the heat carriers when most scattering events conserve momentum
and do not dissipate heat flux. This is the case for two-dimensional or layered
materials at room temperature, or three-dimensional crystals at cryogenic
temperatures. In this work we show that the eigenvectors of the scattering
matrix in the Boltzmann equation define collective phonon excitations, termed
here relaxons. These excitations have well defined relaxation times, directly
related to heat flux dissipation, and provide an exact description of thermal
transport as a kinetic theory of the relaxon gas. We show why Matthiessen's
rule is violated, and construct a procedure for obtaining the mean free paths
and relaxation times of the relaxons. These considerations are general, and
would apply also to other semiclassical transport models, such as the electronic
Boltzmann equation.
For heat transport, they remain relevant even in conventional crystals like silicon, but are of
the utmost importance in the case of two-dimensional materials, where they can revise
by several orders of magnitude the relevant time- and length-scales for thermal
transport in the hydrodynamic regime.
\end{abstract}

\maketitle

\section{Introduction}

The foundations for the theories of lattice thermal transport have been set in place long ago, from the phonon Boltzmann transport equation (BTE) \cite{peierls} to Green-Kubo linear-response theory \cite{green,kubo}.
However, only recently it has become possible, thanks to our increased computational capabilities,
to solve these transport models with high accuracy and without resorting to oversimplifying assumptions \cite{broido:silicon,broido:apl,marzari:prl,fugalloPRB,esfarjani:1,broido:prb80,Luckyanova16112012, cahill:review,donadio:prl,gaughey:md_and_bte,volz:md_and_bte,baowen:nanotubes}.

In particular, the linear BTE can nowadays be solved exactly, using empirical or first-principles interactions, with iterative \cite{sparavigna:nc,sparavigna:prb53,broido:silicon}, variational \cite{ziman,parrot:prb178,srivastava:1976,fugalloPRB} or direct diagonalisation algorithms \cite{guyer:pr1,hardy-ss,chaput:prl}, all of which do not need to simplify the scattering operator with the often-adopted single-mode relaxation time approximation (SMA).
In the SMA, each phonon mode relaxes independently to equilibrium, and it has long been known that this is an incorrect assumption for solids at low temperatures \cite{ziman,callaway:pr}.
Importantly, this approximation fails dramatically in lower dimensions, as first found in graphene \cite{broido:graphene1}, boron nitride and other two-dimensional (2D) materials \cite{nature-mio,broido:isobn,gaughey:phosphorus}, as well as in layered crystals \cite{fugallo-nano}.
The origin of such failure of the SMA has been up to now a matter of debate, with an emerging picture of collective phonon excitations being responsible for heat transfer \cite{fugallo-nano,Comeau-graphene-emd,nature-mio,nature-chen,barbarino-mfp-graphene,spagnoli}, while nevertheless lacking a definition of such excitations.

The microscopic interpretation of thermal transport in the BTE is based on the kinetic theory of gases, used in various contexts since its development in the 19$^{\text{th}}$ century, which relates the thermal conductivity to the velocities and relaxation times of the carriers, with phonons being usually identified as the relevant gas of excitations.
However, as argued below, this identification is incorrect, since only the adoption of the SMA allows for the definition of a time interval (i.e. a lifetime) between heat flux dissipation events taken as phonon scatterings.
Going beyond the SMA, the full, exact solution of the BTE provides the correct thermal conductivity (dramatically improved in 2D materials or at low temperatures), but adds complexity in its interpretation.
In fact, solving exactly the BTE implies abandoning the concept of phonon relaxation time and the description of heat being carried by a gas of phonons.
In other words, phonon lifetimes or phonon mean free paths are no longer relevant quantities to describe thermal transport, since phonons are not the heat carriers anymore.
Yet, the beauty of the SMA description lies in the simplicity of its description of transport.
A natural question then arises: can one define heat carriers, relaxation times, or mean free paths within the exact treatment of the BTE?

In this work, we provide an answer to these questions.
First, we recall why phonon lifetimes are unrelated to heat flux dissipation. 
Then, we define a set of collective excitations, termed relaxons, that diagonalize the scattering matrix.
The BTE is rewritten in the basis of these relaxons; in this representation, each eigenvector represents a collective excitation consisting of a linear combination of out-of-equilibrium phonon populations, and it describes the thermal relaxation of a collective excitation of out-of-equilibrium lattice vibrations.
We show that each relaxon is characterised by a well defined relaxation time; in the case of a homogeneous system at the steady-state each relaxon has also a well defined velocity and mean free path, and the thermal conductivity can be interpreted exactly as a kinetic theory of the relaxon gas.
As a practical example, we compare thermal conductivities in graphene and silicon contrasting the relaxon and the phonon representations, and highlight the profoundly different pictures that emerge.

\section{Approximated Relaxation Times}

We start our derivation by recalling the microscopic description of heat transport given by the linearised phonon BTE \cite{ziman}:
\begin{equation}
\frac{\partial n_{\mu}(\boldsymbol{x},t)}{\partial t} + \boldsymbol{v}_{\mu} \cdot \nabla n_{\mu}(\boldsymbol{x},t) = - \frac{1}{\mathcal{V}} \sum_{\mu'} \Omega_{\mu\mu'} \Delta n_{\mu'}(\boldsymbol{x},t) \; .
\label{phonon_bte_1}
\end{equation}
This equation describes the out-of-equilibrium dynamics of the phonon excitation number $n_{\mu}$ at position $\boldsymbol{x}$ and time $t$, for all possible phonon states $\mu$ (in shorthand notation $\mu\equiv(\boldsymbol{q},s)$, where $\boldsymbol{q}$ varies over the Brillouin-zone and $s$ over the phonon branches). 
Furthermore, $\boldsymbol{v}_{\mu}$ is the phonon group velocity, $\mathcal{V}$ is a normalization volume, $\Omega_{\mu\mu'}$ is the linear phonon scattering operator and $\Delta n_{\mu} = n_{\mu}-\bar{n}_{\mu}$ is the deviation of the phonon distribution from thermal equilibrium, i.e. the Bose--Einstein distribution $\bar{n}_{\mu}(\boldsymbol{x},t) = (e^{\hbar\omega_{\mu}/k_BT(\boldsymbol{x},t)}-1)^{-1}$, with $\omega_{\mu}$ being the phonon frequency and $T(\boldsymbol{x},t)$ the local temperature.
This linear approximation, commonly used in most studies of transport, allows to describe scattering as a linear operator represented by the action of the matrix $\Omega_{\mu\mu'}$ on $\Delta n_{\mu}$: this assumption holds for small deviations from thermal equilibrium and will always be used in the rest of the manuscript.
The scattering matrix appearing in Eq. (\ref{phonon_bte_1}) is in its most general form and describes all possible mechanisms by which a phonon excitation can be transferred from a state $\mu$ to a state $\mu'$.
For the sake of simplicity, we will limit this manuscript to the inclusion of three-phonon processes and isotopic scattering events  \cite{fugalloPRB}, whose expressions are reported for completeness in Appendix A.

For later convenience, it is useful to write the left-hand side of Eq. (\ref{phonon_bte_1}) in terms of the unknown $\Delta n_{\mu}$:
\begin{widetext}
\begin{equation}
\frac{\partial \bar{n}_{\mu}}{\partial T}
 \bigg( \frac{\partial T(\boldsymbol{x},t)}{\partial t} 
+ \boldsymbol{v}_{\mu} \cdot \nabla T(\boldsymbol{x},t) \bigg)
+ \frac{\partial( \Delta n_{\mu}(\boldsymbol{x},t))}{\partial t} + \boldsymbol{v}_{\mu} \cdot \nabla (\Delta n_{\mu}(\boldsymbol{x},t) )
= - \frac{1}{\mathcal{V}} \sum_{\mu'} \Omega_{\mu\mu'} \Delta n_{\mu'}(\boldsymbol{x},t) \; ;
\label{phonon_bte}
\end{equation}
\end{widetext}
where $T$ is the reference temperature at which the BTE has been linearized.
To obtain Eq. (\ref{phonon_bte}), we substituted  $n_{\mu} =  \bar{n}_{\mu} + \Delta n_{\mu}$ in Eq. (\ref{phonon_bte_1}) and used the fact that the Bose--Einstein distribution depends on space and time only through the temperature $T(\boldsymbol{x},t)$.

A closed-form solution of the above equation can be obtained in the SMA, which replaces the scattering operator with its diagonal terms
\begin{align}\label{eq2}
\frac{1}{\mathcal{V}} \sum_{\mu'} \Omega_{\mu\mu'} \Delta n_{\mu'}(\boldsymbol{x},t)   
&\approx   \frac{ \Delta n_{\mu}(\boldsymbol{x},t)   }{\tau^{ \text{SMA}}_{\mu}} \;.
\end{align}
To show that in this simplified diagonal form $\tau_{\mu}^{\text{SMA}}$ represents indeed a relaxation time, let's consider a system at thermal equilibrium ($T(\boldsymbol{x},t) = T$), so that the phonon distribution is $\bar{n}_{\mu}$ everywhere and thus in Eq. (\ref{phonon_bte}) $\nabla (\Delta n_{\mu})=0$, $\frac{\partial T}{\partial t}=0$ and $\nabla T = 0$.
If we excite a single phonon at time $t_0$, its population relaxes back to equilibrium as $\Delta n_{\mu}(t) = (n_{\mu}(t_0) - \bar{n}_{\mu}) e^{-t/\tau_{\mu}^{\text{SMA}}}$, i.e. with a characteristic time $\tau_{\mu}^{\text{SMA}}$.

The thermal conductivity tensor $k^{ij}$ ($i$ and $j$ are cartesian indices) is defined as the ratio between a heat flux $Q^i$ and a static gradient of temperature $(\nabla T)^j$.
Two simplifications apply in this case.
First, a steady-state condition allows to simplify the BTE by setting time derivatives to zero. 
Second, the spatial gradient can be simplified taking $\nabla (\Delta n_{\mu}) = 0$.
This assumption, frequently adopted in literature, holds for a homogeneous perturbation of a bulk crystal (as in our case): if we apply a thermal gradient to a crystal at temperature $T$, the response $\Delta n_{\mu}$ should not depend on the particular position $\boldsymbol{x}$ inside the sample.
Although we will not consider it further here, we note that this assumption cannot be applied when studying systems that break translational invariance, involving e.g. surfaces or pointlike heat sources.
Under these conditions and the SMA, the resulting BTE can be solved analytically and, using the harmonic approximation for the heat flux $Q = \frac{1}{\mathcal{V}}\sum_{\mu} \hbar \omega_{\mu} v_{\mu} \Delta n_{\mu}$ \cite{hardy-flux} and the definition $Q^i=-\sum_jk^{ij}(\nabla T)^j$, the thermal conductivity is given by
\begin{equation}
(k^{ij})^{\text{SMA}} 
= \frac{1}{\mathcal{V}}\sum_{\mu} C_{\mu} v^i_{\mu} (\Lambda^j_{\mu})^{\text{SMA}} \; ,
\label{k_sma}
\end{equation}
where $(\Lambda^j_{\mu})^{\text{SMA}}$ is the component of the phonon mean free path in direction $j$.
This expression can be interpreted as the thermal conductivity of a gas of phonons, each carrying a specific heat $C_{\mu}=\frac{1}{k_BT^2} \bar{n}_{\mu}(\bar{n}_{\mu}+1) (\hbar\omega_{\mu})^2 = \frac{\partial \bar{n}_{\mu}}{\partial T} \hbar \omega_{\mu}$, traveling at velocity $v^i_{\mu}$ and with a mean free path $(\Lambda^j_{\mu})^{\text{SMA}}=v^j_{\mu} \tau_{\mu}^{\text{SMA}}$ before being thermalised by scattering.
Crucially, the definition of phonon lifetime or mean free path cannot be extended beyond the SMA, since the off-diagonal terms of the scattering operator introduce couplings between phonons, and phonon thermalisation stops being governed by an exponential relaxation.

\section{Relaxons}

An exact definition of relaxation times has been formally derived by Hardy \cite{hardy-ss}, as an auxiliary result in his study of second sound.
To recall it, let's first note that the left side of the BTE in Eq. (\ref{phonon_bte}) has a drifting operator diagonal in $\mu$, whereas the right side has a scattering operator (determining scattering time scales) that is non-diagonal.
To identify meaningful scattering times, we proceed with a change of basis that diagonalises the scattering operator while allowing the drifting term to become non-diagonal.
To make more apparent the symmetries within the BTE, we perform the transformations \cite{hardy:symmetrisation,krumhansl:symmetrisation,hardy-ss,chaput:prl}:
\begin{gather}
\tilde{\Omega}_{\mu\mu'} =  \Omega_{\mu\mu'}   \sqrt{ \frac{\bar{n}_{\mu'} (\bar{n}_{\mu'}+1)}{\bar{n}_{\mu} (\bar{n}_{\mu}+1) }} \;, \text{ and}
\label{n_symmetrisation} \\
\Delta \tilde{n}_{\mu} =  (\bar{n}_{\mu} (\bar{n}_{\mu}+1) )^{-\frac{1}{2}} \Delta n_{\mu}\;.
\label{n2_symm}
\end{gather}
These transformations are introduced to scale quantities appearing in the BTE in such a way that $\tilde{\Omega}_{\mu\mu'}=\tilde{\Omega}_{\mu'\mu}$ (the matrix $\Omega$ does not obey this symmetry; see Appendix A for a detailed explanation).
We note that sometimes these transformations appear in literature in the form of hyperbolic sines, by means of the identity $\sinh(\frac{\hbar\omega_{\mu}}{2k_BT}) = \frac{1}{2\sqrt{\bar{n}_{\mu} (\bar{n}_{\mu}+1)}}$.
Since $\tilde{\Omega}$ is a real symmetric matrix, it can be diagonalized, 
giving eigenvectors $\theta_{\mu}^{\alpha}$ and real eigenvalues $\frac{1}{\tau_{\alpha}}$ such that
\begin{equation}
\frac{1}{\mathcal{V}} \sum_{\mu'} \tilde{\Omega}_{\mu\mu'} \theta_{\mu'}^{\alpha}  =  \frac{1}{\tau_{\alpha}} \theta_{\mu}^{\alpha} \; ,
\label{diagonai}\end{equation}
where $\alpha$ is the eigenvalue index. 
In passing, we define the scalar product $\big< \alpha \big| \alpha' \big>  \equiv \frac{1}{\mathcal{V}} \sum_{\mu} \theta_{\mu}^{\alpha} \theta_{\mu}^{\alpha'}$ that allows to define the orthonormalization condition for the eigenvectors ( $\big< \alpha \big| \alpha' \big>=\delta_{\alpha\alpha'}$ ) which will be helpful in the next algebraic operations.
It can be shown \cite{hardy:symmetrisation,hardy-ss} that $\tilde{\Omega}$ is positive-semidefinite, i.e. $\frac{1}{\tau_{\alpha}}\ge0$ $\forall \alpha$, and that its eigenvectors are either even or odd, i.e. $\theta^{\alpha}_{\mu} = \pm \theta^{\alpha}_{-\mu}$, where $-\mu=(-\boldsymbol{q},s)$ \cite{hardy-ss}.
Little else is known on the eigenvalue spectrum of Eq. (\ref{diagonai}), which therefore has to be characterized numerically.
In contrast with Refs. \cite{hardy-ss,guyer:pr1}, we remark that the Bose--Einstein distribution is not an eigenvector with zero eigenvalue: the scattering operator acts only on the deviation from equilibrium $\Delta n_{\mu}$, therefore thermal equilibrium ($\Delta n_{\mu} = 0$) is a stationary solution ($\sum_{\mu'} \Omega_{\mu\mu'} \Delta n_{\mu}=0$) because it's an algebraically trivial solution.
However, the Bose--Einstein distribution allows the introduction of a vector of unitary length
\begin{equation}
\theta^0_{\mu} = 
\frac{  \sqrt{\bar{n}_{\mu} (\bar{n}_{\mu}+1) } \hbar \omega_{\mu}   }  { \sqrt{k_B T^2 C} } \;,
\end{equation}
where $C = \frac{1}{\mathcal{V}}\sum_{\mu}C_{\mu}$, describing the increase of temperature.
This vector is constructed as the linear deviation from equilibrium of $\bar{n}_{\mu}(T+\delta T)$, transformed using Eq. (\ref{n2_symm}) and normalized to one.
Note that $\theta^0_{\mu}$ is not an eigenvector and doesn't have to be orthogonal to other eigenstates $\alpha$.

Any response $\Delta \tilde{n}_{\mu}$ can be written as a linear combination of the $\theta_{\mu}^{\alpha}$ eigenvectors \cite{hardy-ss}
\begin{equation}
\Delta \tilde{n}_{\mu}(\boldsymbol{x},t) = \sum_{\alpha} f_{\alpha}(\boldsymbol{x},t) \theta^{\alpha}_{\mu}\;,
\label{basis_change}
\end{equation}
and the BTE can be written in this $\theta^{\alpha}$ basis (to this aim, substitute Eq. (\ref{basis_change}) in (\ref{phonon_bte}) and take the scalar product of the equation with a generic eigenvector $\alpha'$), becoming:
\begin{widetext}
\begin{equation}
\sqrt{ \frac{C}{k_B T^2} }  \bigg( \frac{  \partial T(\boldsymbol{x},t) }{\partial t} \big< 0 \big| \alpha \big>  + \nabla T(\boldsymbol{x},t) \cdot \boldsymbol{V}_{\alpha}  \bigg)
+ \frac{\partial f_{\alpha}(\boldsymbol{x},t)}{\partial t} 
+ \sum_{\alpha'} \boldsymbol{V}_{\alpha\alpha'} \cdot \nabla f_{\alpha'}(\boldsymbol{x},t) 
= -  \frac{ f_{\alpha}(\boldsymbol{x},t) }{\tau_{\alpha}}   \;,
\label{relaxon_bte}
\end{equation}
\end{widetext}
where $\boldsymbol{V}_{\alpha\alpha'} = \frac{1}{\mathcal{V}}\sum_{\mu} \theta_{\mu}^{\alpha} \boldsymbol{v}_{\mu} \theta_{\mu}^{\alpha'} \equiv \big< \alpha \big| v \big| \alpha' \big>$ and $\boldsymbol{V}_{\alpha} = \boldsymbol{V}_{0\alpha} = \big< 0\big|v\big|\alpha\big>$. 
$\boldsymbol{V}_{\alpha\alpha'}$ derives from the action of the diffusion operator on the deviation from equilibrium, while $\boldsymbol{V}_{\alpha}$ derives from the action of the diffusion operator on the equilibrium distribution.

The physical picture encoded in Eq. (\ref{relaxon_bte}) underlines one of the key statements of this work: by diagonalising the scattering operator, the information about the characteristic relaxation time of the thermal excitations is now given by the eigenvalues $\frac{1}{\tau_{\alpha}}$.
The eigenvectors $\theta^{\alpha}_{\mu}$ for which this picture emerges represent collective excitations which we call here \emph{relaxons}.
Each relaxon represents a distribution of phonon excitation numbers (a wave packet), describing how the phonon distribution is relaxing to equilibrium.
The coefficients $f_{\alpha}$ are the relaxon occupation numbers, which are determined by the BTE in the out-of-equilibrium state, and that at equilibrium are all 0, so that the deviation from equilibrium $\Delta n_{\mu}$ vanishes.

\begin{figure}
\centering
\includegraphics{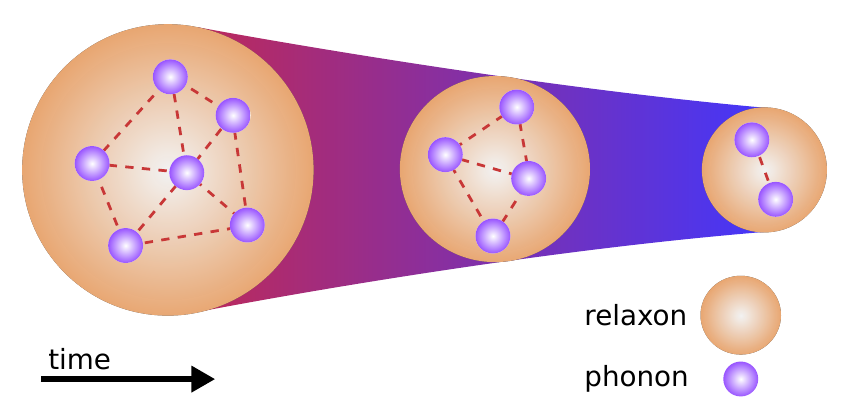}
\caption{
Schematic illustration of the equilibration of lattice vibrations after a thermal excitation. Each relaxon consists of a linear combination of phonons, which interact through scattering events among themselves, but are decoupled from phonons belonging to different relaxons. The relaxon decays exponentially to equilibrium, where it disappears at a rate determined by its relaxation time. 
}
\label{fig1}
\end{figure}

The name relaxon is easily justified by considering a system at thermal equilibrium, so that $\frac{\partial T}{\partial t} = 0$, $\nabla T = 0$ and, in terms of relaxons, all states are empty everywhere, so that $\nabla f_{\alpha}=0$ $\forall \alpha$ in Eq. (\ref{relaxon_bte}).
If we excite a single relaxon $\alpha$ at time $t_0$, its occupation will relax back to equilibrium as $f_{\alpha}(t) = f_{\alpha}(t_0) e^{ - t / \tau_{\alpha}}$, therefore endowing $\tau_{\alpha}$ with the meaning of a relaxation time.
Although the theory allows for zero eigenvalues, we will find in our examples only strictly positive relaxation times, so that all relaxons decay to zero for $t\to\infty$.
Using Eq. (\ref{basis_change}), one can show instead that phonon populations do not have well-defined relaxation times: since each phonon population decays as a linear combination of relaxon processes $\Delta n_{\mu}(t) = \sum_{\alpha} f_{\alpha}(t_0) \theta_{\mu}^{\alpha} e^{-t/\tau_{\alpha}}$, the characteristic time for the decay depends on the initial conditions of the thermal excitation and can even display damped oscillations.
Conversely, let's suppose to excite only one phonon mode $\mu$ at time $t_0$.
This initial state can be decomposed as the sum of different relaxons, each evolving with a different relaxation time.
Therefore, at a subsequent time $t$ one will also observe that new phonon modes $\mu'\ne\mu$ have been excited out of thermal equilibrium.
In Fig. \ref{fig1} we graphically illustrate our interpretation: each relaxon is a collective excitation of phonons, which interact through scattering events among themselves, but are decoupled from phonons belonging to different relaxons; owing to their positive relaxation time, relaxons disappear at long times, allowing the system to reestablish equilibrium. 

Velocities appear in Eq. (\ref{relaxon_bte}) with a matrix $\boldsymbol{V}_{\alpha\alpha'}$ coupling different relaxons, since it is the phonon basis that diagonalises the drifting operator; therefore, one cannot always identify a  relaxon velocity.
However, if we suppose to work in an infinite crystal at temperature $T$ and apply a temperature perturbation homogeneously to the entire crystal, the response of the system is constant throughout the space and we can set $\nabla f_{\alpha}=0, \forall \alpha$.
Therefore, the BTE simplifies to
\begin{equation}
\frac{\partial f_{\alpha}(t)}{\partial t} 
+ \sqrt{\frac{C}{k_B T^2}} \bigg( \frac{ \partial T }{\partial t} \big<0 \big| \alpha \big> + \nabla T \cdot \boldsymbol{V}_{\alpha} \bigg)
= - \frac{ f_{\alpha}(t) }{ \tau_{\alpha} } \;.
\label{homo_relaxon_bte}
\end{equation}
Both the drifting and the collision operator are now diagonal in $\alpha$, and $\boldsymbol{V}_{\alpha}$ identifies a well-defined relaxon velocity.

Let's simplify the problem further and consider steady state.
In this case, time derivatives are set to zero in Eq. (\ref{homo_relaxon_bte}) and one can, for small deviations from equilibrium, search for linear solutions of the form $f_{\alpha} = \sum_i f_{\alpha}^i \nabla_i T$, where $i$ is a cartesian direction.
The BTE reduces to
\begin{equation}
\sqrt{\frac{C}{k_B T^2}} V^i_{\alpha} = - \frac{ f^i_{\alpha} }{ \tau_{\alpha} } \;,
\label{steady_state_relaxon}\end{equation}
whose solution for $f_{\alpha}^i$ is trivial.
Using the relation between phonons and relaxon occupation numbers $\Delta n_{\mu} = \sqrt{\bar{n}_{\mu}(\bar{n}_{\mu}+1)} (\sum_{i\alpha} \nabla_i T f^i_{\alpha} \theta_{\mu}^{\alpha} )$, we obtain the thermal conductivity
\begin{align}
k^{ij} &= \frac{-1}{\mathcal{V} \nabla_i T } \sum_{\mu} \hbar \omega_{\mu} v_{\mu}^j \Delta n_{\mu} = 
-\sum_{\alpha} f_{\alpha}^i \sqrt{k_B T^2 C} V_{\alpha}^j  \nonumber \\
&= \sum_{\alpha} C V_{\alpha}^i V_{\alpha}^j \tau_{\alpha}
=\sum_{\alpha} C V_{\alpha}^i \Lambda_{\alpha}^j \; ,
\label{k_relaxon}
\end{align}
where we introduced the relaxon mean free path $\Lambda_{\alpha}$ ($\Lambda^j_{\alpha}$ is the component of $\Lambda_{\alpha}$ in direction $j$).
Therefore, the exact thermal conductivity in Eq. (\ref{k_relaxon}) is expressed in the framework of the kinetic theory of gases, and thermal transport can be thought as a flux of relaxons, each carrying a specific heat $C$, traveling at velocity $V_{\alpha}$ for an average distance of $\Lambda_{\alpha}$ before thermalisation occurs.

At variance with the phonon picture, where each phonon participates to thermal conductivity with a mode specific heat $C_{\mu}$, all relaxons contribute with the same specific heat of the crystal $C$.
Mathematically, the phonon-mode specific heat is moved in the vector $\theta_{\mu}^0$ (note that $(\theta_{\mu}^0)^2 = \frac{C_{\mu}}{C}$ ) and thus is included into the relaxon velocity $V_{\alpha} = \big< 0 \big| v \big| \alpha\big>$.
To physically interpret this difference we recall that, from a thermodynamic point of view, the quantity $C \delta T$ is the energy needed by the system to change temperature by $\delta T$.
To observe such temperature change, all phonon modes must simultaneously change their occupation number according to the collective excitation $\theta_{\mu}^0$.
The quantity $C_{\mu} \delta T$ is the decomposition of such energy change in terms of each phonon mode. 
However, as explained before, one cannot suppose to excite a single phonon mode and bring it to a higher temperature without affecting the rest of the phonon ensemble: phonon scattering would redistribute the energy excess of such mode to the rest of the system. 
Only a collective excitation of phonons ($\theta_{\mu}^0$) leads to a temperature change and the total energy cost for increasing temperature is necessarily associated with $C$; thus, 
from a thermodynamic point of view, one could state that the mode specific heat $C_{\mu}$ does not have 
a well-defined meaning.

It's worth to point out the role played by the parity of relaxons.
The quantity $\boldsymbol{V}_{\alpha} = \big<0 \big| \boldsymbol{v} \big|\alpha\big>$ involves the odd function $\boldsymbol{v}_{\mu}$ ($-\boldsymbol{v}_{\mu}=\boldsymbol{v}_{-\mu}$) and the even function $\theta^0_{\mu}$ (owing to $\omega_{\mu}=\omega_{-\mu}$).
Therefore, relaxon velocities $\boldsymbol{V}_{\alpha}$ are different from zero only for odd relaxons $\alpha$.
Consequently, Eq. (\ref{steady_state_relaxon}) implies also that only odd relaxons are excited in the steady state condition, and thus contribute to heat flux, while even relaxons have zero occupation number.
The role of parity is reversed for determining the energy of the system, since the change from equilibrium energy $\Delta E$ is
\begin{align}
\Delta E &= \frac{1}{\mathcal{V}} \sum_{\mu}  \hbar \omega_{\mu}\Delta n_{\mu}
=  \frac{1}{\mathcal{V}} \sum_{\mu} \hbar \omega_{\mu}   \sqrt{\bar{n}_{\mu} (\bar{n}_{\mu}+1)  }  \Delta \tilde{n}_{\mu}   \nonumber \\
&=  \frac{1}{\mathcal{V}} \sum_{\mu}   \theta^0_{\mu}   \sqrt{k_BT^2C}\sum_{\alpha} f_{\alpha} \theta^{\alpha}_{\mu} \nonumber \\
&=  \sqrt{k_BT^2C} \sum_{\alpha} f_{\alpha} \big<0\big|\alpha\big> \;.
\end{align}
In this case, even relaxons have a non-zero coefficient $\big<0\big|\alpha\big>$ and contribute to an energy change, but for odd relaxons $\big< 0\big|\alpha\big>=0$ and thus do not change energy.
We can thus deduce that at the steady state defined by Eq. (\ref{homo_relaxon_bte}), where only odd relaxons are excited, the energy of the system is conserved.

\section{Graphene}

\begin{figure}
\centering
\includegraphics{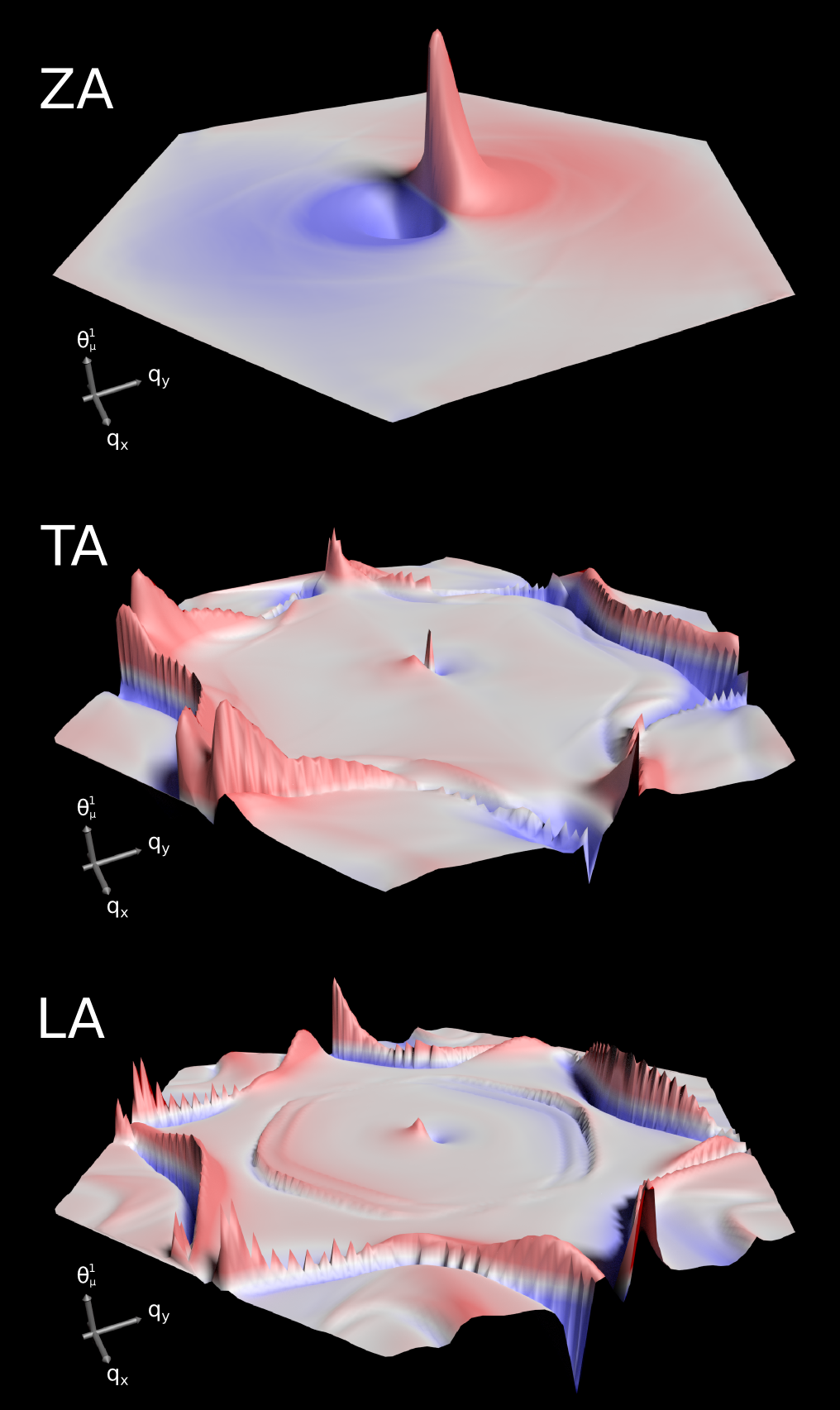}
\caption{
Representation of the relaxon $\theta_{\mu}^{\alpha}$ with the longest relaxation time ($\alpha$=1) in graphene at room temperature as a function of the phonon index $\mu=(\boldsymbol{q},s)$, where we choose $s$ to be the out-of-plane/transverse/longitudinal acoustic mode (ZA, TA and LA respectively).
We recall that the relaxon is a difference in phonon populations with respect to thermal equilibrium: overpopulated modes are colored in red, depopulated ones are in blue.
The fine structure of the ridges is a numerical artifact due to discrete Brillouin zone sampling.
}
\label{fig2}
\end{figure}

As a first numerical example supporting these conclusions, we study relaxons in graphene, the material with the highest known thermal conductivity \cite{Balandin:nanoletter}, and contrast the phonon and the relaxon pictures at 300K.
Due to its symmetry, graphene's $k^{ij}$ tensor is diagonal and, since $k^{xx}=k^{yy}$, it has only one independent component (verified also numerically); therefore in the following we will drop cartesian indices and compute quantities numerically along the zig-zag direction.
To proceed, we calculate harmonic and anharmonic force constants using density-functional perturbation theory \cite{baroni:rev,giannozzi:prb43,debe:prl,paulatto,lazzeri:anharmonic,baroni:prl-dfpt,gonze:2np1} as implemented in the Quantum-ESPRESSO distribution \cite{qe} and construct the scattering matrix using 3-phonon and isotope-phonon interactions.
The diagonalisation of Eq. (\ref{diagonai}) provides all the relaxon eigenvectors $\theta_{\mu}^{\alpha}$; each of them represents, at fixed relaxon index $\alpha$, a difference in phonon populations with respect to thermal equilibrium (provided that is back-transformed using Eq. (\ref{n2_symm})).
Notably, only few $\theta_{\mu}^{\alpha}$ have large relaxation times; for example, the longest-lived relaxon ($\alpha$=1) is plotted in Fig. \ref{fig2} as a function of the phonon index $\mu=(\boldsymbol{q},s)$.
The first three ($s=1,2,3$) branches are shown, corresponding to the out-of-plane, transverse or longitudinal acoustic phonons  (ZA, TA and LA respectively).
This particular relaxon induces a population difference for the ZA branch mainly located close to the Brillouin zone center, whereas TA and LA modes are altered throughout the Brillouin zone.
The variations of optical modes ($s=4,5,6$ not shown) are an order of magnitude smaller.
The complex landscape drawn by these phonon distributions reflects the fact that out-of-equilibrium lattice properties cannot be described in terms of single phonon properties, as the action of scattering tightly couples phonons of any wavevector and branch.

\begin{figure}
\centering
\includegraphics{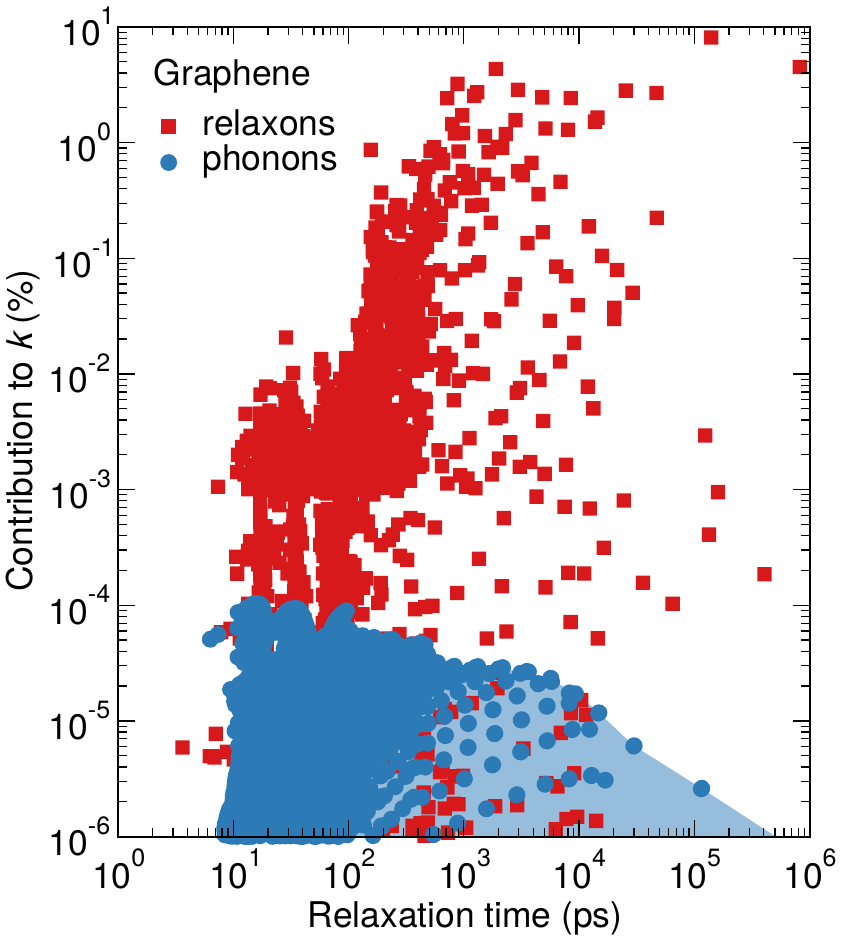}
\caption{
Relaxation times and their contribution to the thermal conductivity of graphene at room temperature, considering relaxons or phonons as heat carriers.
Relaxons tend to be longer lived than single phonon excitations, with large contributions to thermal conductivity coming from excitations with relaxation time larger than 10$^3$ps, whereas phonons have lifetimes mainly in the range 10-100 ps.
The shaded area is a guide to the eyes to stress that phonos form a continuous spectrum, while relaxons are discretized: thermal conductivity can be accurately described using a small number of relaxons.
}
\label{fig3}
\end{figure}

We analyse the entire phonon and relaxon spectrum in Fig. \ref{fig3}, where the contributions to the SMA or the exact thermal conductivities are plotted as a function of the relaxon or phonon relaxation times.
The thermal conductivities computed in the two pictures differ significantly in graphene (in this work we compute 3894 W/mK with the exact BTE against 495 W/mK with the SMA), hence, to have more comparable quantities, we plot the percentage contribution to thermal conductivity.
We first note that the spectrum of phonon lifetimes (and phonon velocities and mean free paths) is continuous, with a divergence $\tau_{\mu}\to\infty$ for acoustic ZA phonons at the $\Gamma$ point \cite{boniniNANOL}.
This divergence cannot be accurately described with a finite mesh of points sampling the Brillouin-zone (in our case, a full mesh of 128$\times$128 points), resulting in a sparse tail of long-lived phonons on the right side of Fig. \ref{fig3}, whose contribution to $k^{\text{SMA}}$ is negligible \cite{boniniNANOL}.
Instead, relaxation times for relaxons are discrete and sparse, in particular in the region of large values, so that only a small number of relaxons is sufficient to describe thermal transport with high accuracy.
This observation is robust with respect to Brillouin Zone sampling: an improvement in the integration mesh has new phonon modes appear in the long-lifetime region; instead the longest relaxon relaxation times converge - from above - to the discretized values shown in figure.
On average, relaxon relaxation times are larger by at least two orders of magnitude with respect to phonon lifetimes.
The large difference between the time scales of phonons and relaxons appears because a single phonon scattering cannot thermalize the system \cite{ziman}, as instead implied by the SMA.
Therefore, while phonons scatter at timescales of about 10-100 ps, heat flux is dissipated by relaxons within nano- and micro-second timescales.

\begin{figure}
\centering
\includegraphics{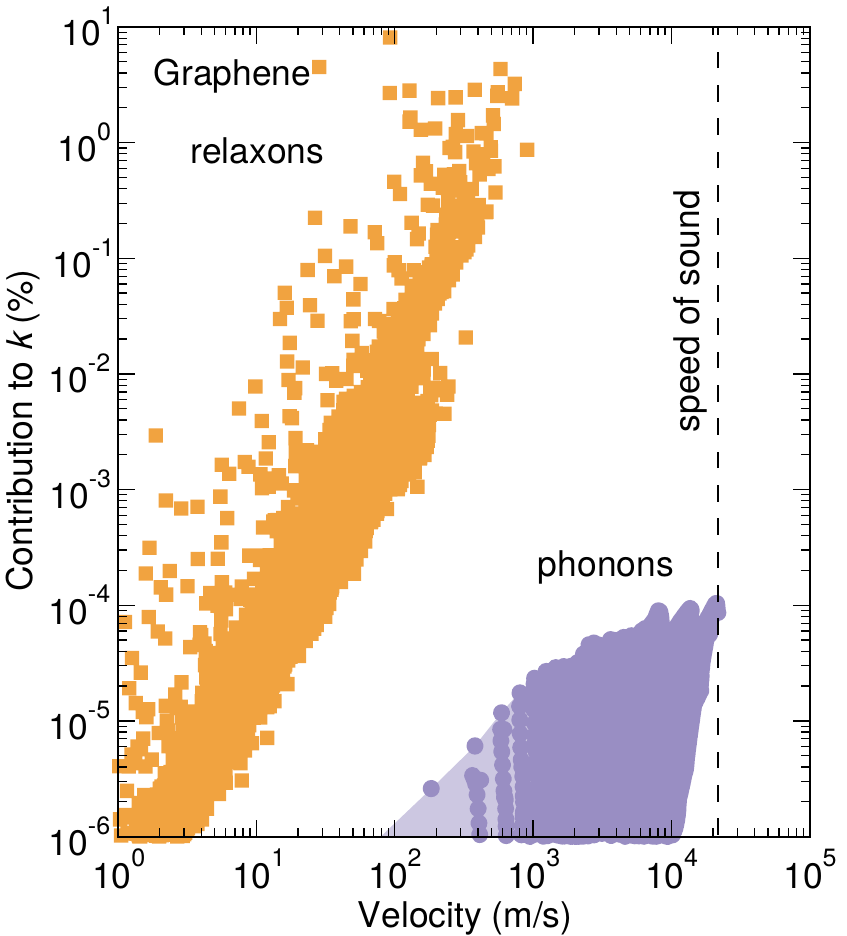}
\caption{Comparison of relaxon and phonon velocities and their contributions to the room temperature thermal conductivity of graphene. 
When approximating phonons as heat carriers, the velocity scale of thermal transport is set by the speed of sound (about 20 km/s for longitudinal acoustic phonons in graphene). 
Instead, relaxon velocities are at least an order of magnitude smaller, illustrating how much the phonon scattering slows down the heat flux.
}
\label{fig4}
\end{figure}

Before analysing velocities, we note that the sign of $V_{\alpha}$ is arbitrary, since both $\theta^{\alpha}_{\mu}$ and $-\theta^{\alpha}_{\mu}$ are relaxon eigenvectors.
As a convention, we select the sign of odd eigenvectors such that $V_{\alpha}$ is non-negative (and so also $\Lambda_{\alpha}$), noting that in any case the contribution to $k$ would be positive (as $V_{\alpha}^2$).
Phonon velocities can also assume both signs: in figure we plot their absolute values.
Fig. \ref{fig4} reveals that the velocities of relaxons are much smaller than those of phonons: while the scale of phonon velocities is set by the speed of sound (the group velocity of the longitudinal acoustic phonon is about 20 km/s), relaxons are slower by two orders of magnitude, indicating that heat is transferred through the material at 0.1-1 km/s.

\begin{figure}
\centering
\includegraphics{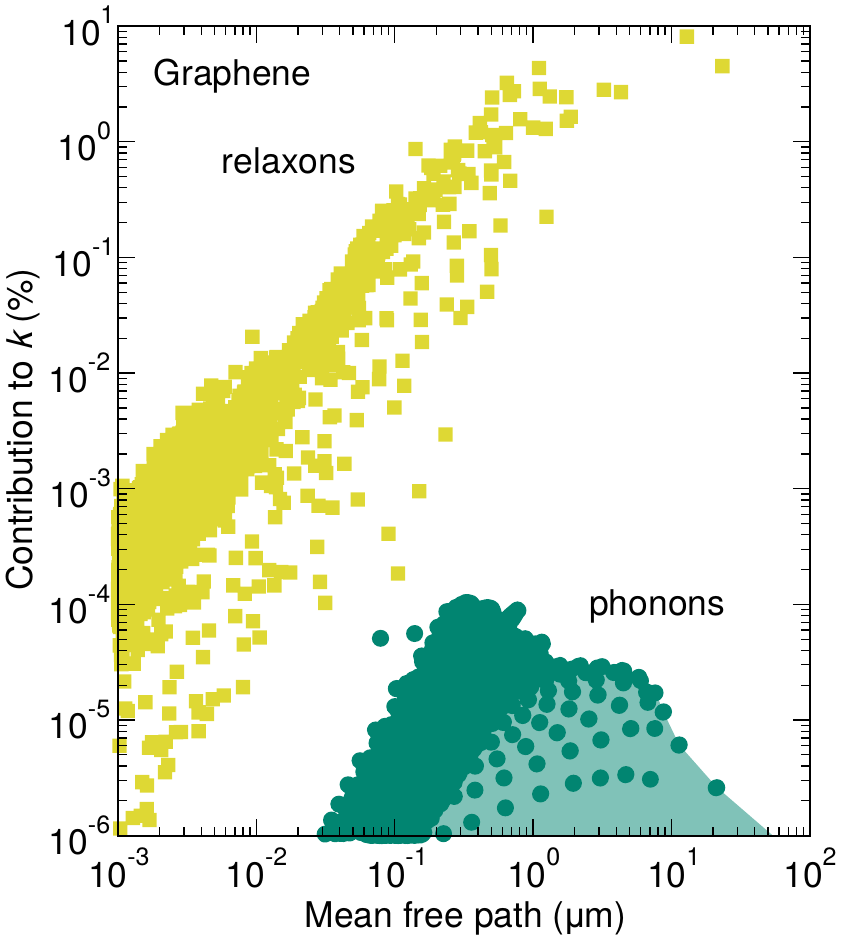}
\caption{Analysis of phonon and relaxon mean free paths and their contributions to the thermal conductivity of graphene at room temperature. The spectrum of phonon mean free paths is peaked in the submicron region, whereas relaxon mean free paths are skewed to values larger than 1$\mu$m. The two largest relaxon mean free paths, illustrating the maximum distance that heat flux can propagate inside the material before decaying, lie between 10 and 100$\mu$m.}
\label{fig5}
\end{figure}

Finally, we show the relaxon mean free paths in Fig. \ref{fig5} (projected along the transport direction).
As other first-principles studies reported \cite{esfarjani:1}, phonon mean free paths for graphene are distributed in the 0.1-1\textmu m region \cite{fugallo-nano}; this is confirmed here.
For relaxons, most contributions to thermal conductivity come with mean free paths above 0.1 \textmu m, the longest and most important contributions having mean free paths up to tens of \textmu m.
The contribution to $k$ is roughly monotonic with the mean free path, and the large increase in $\tau_{\alpha}$ is partly compensated by the decreased $V_{\alpha}$.
The saturation of relaxons' mean free paths at tens of \textmu m appears to be in contrast with recent estimates for saturation lengths of 100 \textmu m \cite{fugallo-nano} or longer \cite{doi:10.1021/acs.nanolett.5b02403}; we will comment this discrepancy after discussing the next example.

\section{Silicon}

\begin{figure}
\centering
\includegraphics{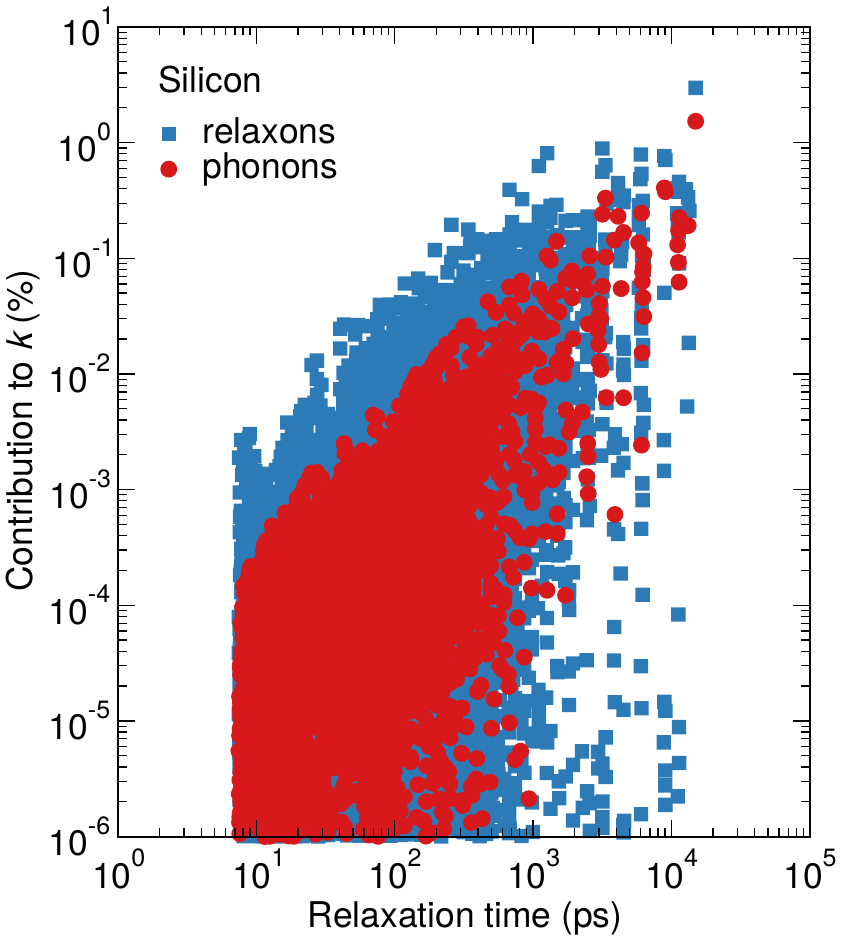}
\caption{Same as in Fig. \ref{fig3} this time for silicon at room temperature, the spectrum of relaxation times and their contribution to thermal conductivity, considering relaxons or phonons as heat carriers.
In this material, the relaxation times estimated with the SMA are relatively close to the exact relaxation times of the system, with contributions ranging from a few picoseconds up to approximately 10$^4$ps.
}
\label{fig3_si}
\end{figure}

\begin{figure}
\centering
\includegraphics{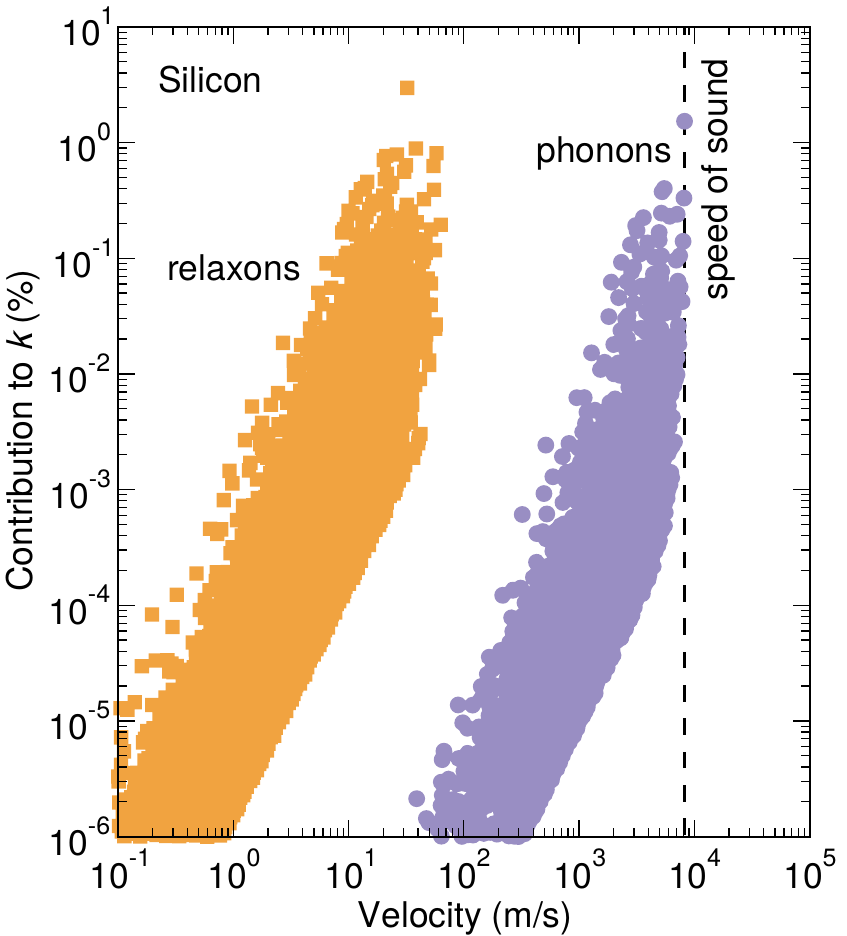}
\caption{Same as in Fig. \ref{fig4} but for silicon, the comparison of relaxon and phonon velocities and their contributions to room temperature thermal conductivity.
Similar to the case of graphene, the velocity of the SMA description is set by the velocity of phonons, with long-wavelength acoustic modes (those with the highest velocities) giving the largest contributions. Relaxons velocities are instead smaller by about two orders of magnitude, indicating that the heat flux is slowed down by the action of phonon scattering.}
\label{fig4_si}
\end{figure}

\begin{figure}
\centering
\includegraphics{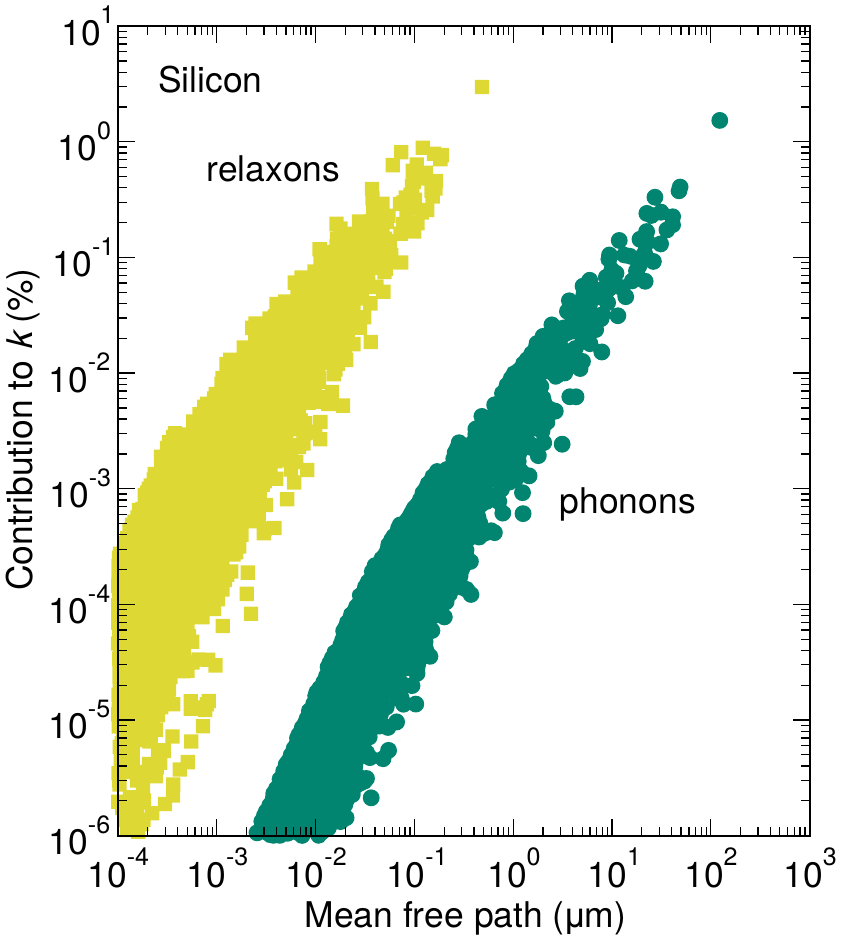}
\caption{Analysis of phonon and relaxon mean free paths and their contributions to the thermal conductivity of silicon at room temperature. The difference between velocities is carried over to this spectrum, so that relaxon mean free paths are shifted to smaller values with respect to phonons. While our estimates of mean free paths for phonons extend up to hundreds of micrometers, relaxon mean free paths barely reach the micrometer scale.}
\label{fig5_si}
\end{figure}


As a further test for relaxons, let us now turn our attention to silicon and examine its thermal transport properties at room temperature.
The thermal conductivity tensor in silicon is diagonal and the three cartesian directions are equivalent; we therefore only consider transport properties along the (100) direction.
At variance with graphene, here the SMA introduces an appealingly small error: in our calculations we find 138 W/mK instead of 141 W/mK for the exact solution; these estimates are in line with previous first-principles studies \cite{broido:apl,marzari:prl,esfarjani:1}.
The small difference between the two pictures is somehow replicated in their relaxation times, reported in Fig. \ref{fig3_si}.
The time-scales covered by phonons and relaxons covers approximately the same range of values, except for one relaxon, not shown in the graph, that has relaxation time of $2\cdot 10^5$ps but negligible contribution to thermal conductivity (10$^{-9}$\%).
However, one can note that the two distributions of values do not perfectly overlap: even if the scattering matrix is diagonally dominant, there are anyway small non-zero out-of-diagonal matrix elements that introduce deviations from the SMA.


It is enlightening to analyze the different velocity scales set by the two pictures, depicted in Fig. \ref{fig4_si}.
Once again, most of the contributions to SMA thermal transport come from phonons with velocities close to the speed of sound, which in silicon is approximately 8 km/s.
However, relaxon velocities are two orders of magnitude smaller than this limiting value, reaching merely 60 m/s.
Despite the fact that the instantaneous velocity of a lattice vibration is determined by the phonon dispersion, the velocity at which the heat flux propagates can be much different: the scattering between phonons slows it down. 

The mean free paths for relaxons and phonons in silicon are compared in Fig. \ref{fig5_si}.
The vast difference originating from the velocities is carried over, so that while mean free paths of phonons extend up to 100\textmu m, in agreement with other first-principles studies \cite{garg2,lindsay2016,esfarjani:1}, relaxons travel for a distance two orders of magnitude smaller than phonons.
It seems therefore puzzling that the two pictures give such large differences in the estimate of velocities and mean free paths, despite the fact that thermal conductivities are essentially identical.
To explain this discrepancy, let's compare Eqs. (\ref{k_sma}) and (\ref{k_relaxon}) for thermal conductivity and recall that specific heat is constant for relaxons and mode-specific for phonons.
Also the SMA conductivity can be written in a form with constant specific heat for each phonon, provided that we rescale velocities as $v_{\mu} \to \sqrt{\frac{C_{\mu}}{\mathcal{V}C}} v_{\mu}$ (consequently, also the mean free path is scaled $\Lambda_{\mu}^{\text{SMA}} = v_{\mu} \tau_{\mu}^{\text{SMA}} \to \sqrt{\frac{C_{\mu}}{\mathcal{V}C}} \Lambda_{\mu}^{\text{SMA}}$).
After this transformation, velocities and mean free paths of phonons and relaxons in silicon are again within the same order of magnitude, although residual discrepancies still persist (see Appendix B for the spectra after rescaling).
Therefore, the differences observed in silicon between the two pictures arise mainly from the different interpretation of specific heat.

We note that other experimental and theoretical efforts have estimated heat mean free paths in silicon \cite{silicon:mfp1,silicon:mfp2,silicon:mfp3,esfarjani:1,garg2,lindsay2016}, obtaining values that are comparable to those of the phonon mean free paths, and different from the relaxon mean free paths presented here.
It is important to stress, though, that at least one of these two assumptions were used: first, that results can be interpreted considering phonons as the heat carriers; and second, that surface or grain boundary scattering can be exploited as a tool to estimate heat mean free paths.
In the present work, we discussed already at length the limitations of the former assumption;
as for the latter one, we cannot compare our results with studies that rely on surface scattering, since the data presented here pertain to a homogeneous bulk crystal.
It is nevertheless possible to drop the bulk condition and solve the BTE in presence of surfaces, reconciling
the different pictures; such discussion goes out of the scope of the present work and will be presented in an upcoming study.

\section{Further Properties}

\begin{figure*}
\centering
\includegraphics{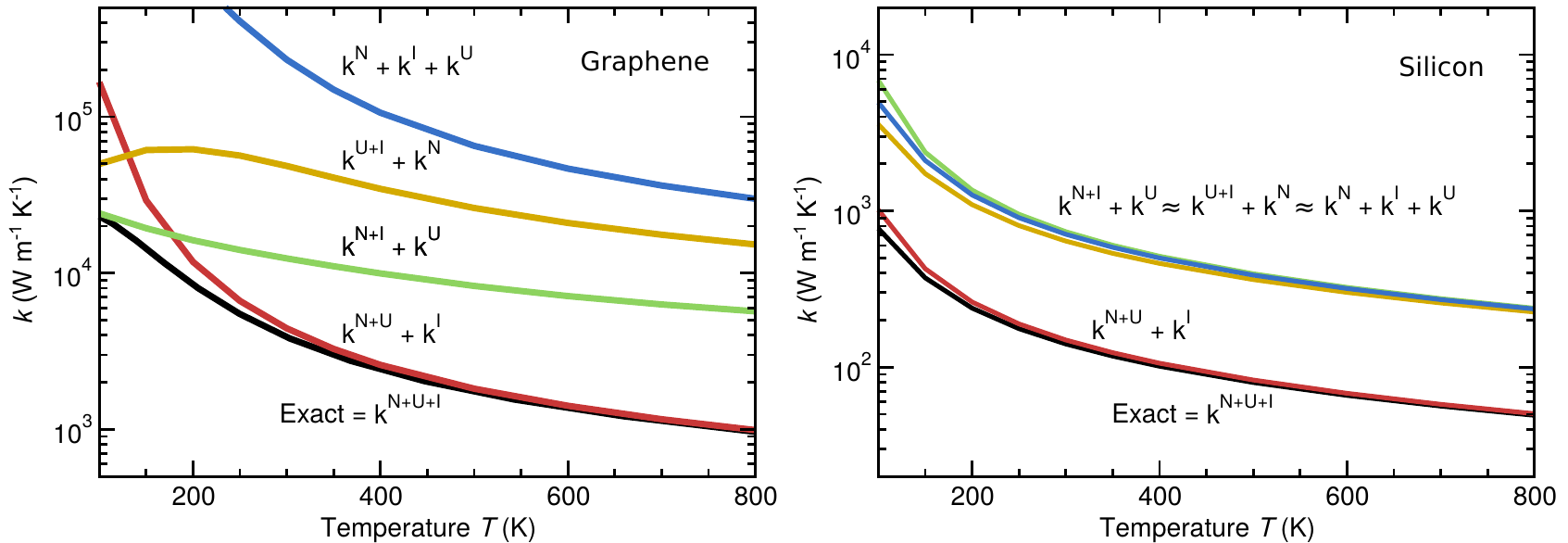}
\caption{Study of the failure of Matthiessen rule in graphene and silicon. The total thermal conductivity (black line) is compared to conductivities computed through various Matthiessen sums, where the total sum is given by the sum of the reciprocals (by $k^N+k^U+k^I$ we mean $k^{-1} = (k^N)^{-1} + (k^U)^{-1}+ (k^I)^{-1}$). Due to the correlation between scattering events, there is no decomposition for which the Matthiessen rule is obeyed at all temperatures. The only curve that approaches the exact thermal conductivity is the decomposition ($k^{N+U}+k^I$), just when the effect of isotopic scattering is negligible.}
\label{fig6}
\end{figure*}

A widely held assumption that is also violated by the exact BTE is the Matthiessen rule,
which states that the total thermal resistivity (i.e. $\frac{1}{k}$) is the sum of the resistivities of each independent scattering mechanism; however, the Matthiessen rule is an approximation \cite{ziman} relying on the possibility of exactly decoupling scattering mechanisms.
To probe numerically this violation, we computed the resistivities of normal, Umklapp and isotopic processes, or any combination of these, and combined them according to Matthiessen rule.
In Fig. \ref{fig6} we show that, regardless of any particular decomposition, the conductivity obtained by imposing the Matthiessen sum deviates significantly from the exact conductivity.
The only case in which a decomposition reproduces the exact result is when the effect of a separated mechanism is negligible; for the case shown in figure, one can sum separately the resistance due to isotopes only at high temperatures, when it's small.
Finally, one can prove that the total thermal conductivity is always smaller or equal to the Matthiessen sum (see \cite{ziman} or Appendix C); this is also verified in our calculations.
Moreover, it is not possible to distinguish the contribution to a relaxon relaxation time due to each scattering mechanism, at variance with the case of a phonon lifetime in the SMA.
This is because it would require that the eigenvalues of the sum of two matrices were the sum of eigenvalues of two matrices - clearly not the case when the scattering matrix is not diagonal.

As an added benefit, the direct diagonalisation of the scattering matrix brings clear insight into the numerical stability of current methods used to solve the BTE.
In particular, we show in Appendix D that the iterative method \cite{sparavigna:nc,sparavigna:prb53,broido:silicon}, often used to study 2D materials, is numerically unstable for graphene at room temperature, due to the dominant contribution of the out-of-diagonal terms in the scattering matrix (this is exactly the case when the relaxon picture differs significantly from the phonon picture).

\section{Conclusions}

\begin{table}
\begin{tabular}{ p{1.7cm}  p{3.2cm} p{3.2cm} }
     & Phonon & Relaxon \\
\hline
Definition & Eigenstate of harmonic Hamiltonian & Eigenstate of collision matrix \\
\hline
  Physical meaning & Collective excitation of atomic displacements & Collective excitation of phonon populations \\
& Quantum of vibrational energy &Elementary carrier of heat \\
\hline
  Exact quantities & Lifetime, velocity and mean free path of the vibration &  Relaxation time, velocity and mean free path of the heat carrier \\
  & Quasiparticle (energy, wavevector, dispersion relations) & No dispersion relations \\
\hline
Thermal conductivity & Only obtained as solution of the BTE & Obtained as a kinetic theory of the relaxon gas \\
\hline
\end{tabular}
\caption{A comparison of the main properties of phonons and relaxons.} \label{table1}
\end{table}

In summary, we have shown that by choosing the eigenvectors of the scattering matrix as a basis, the linear BTE can be greatly simplified.
These eigenvectors are collective excitations of phonon populations, termed relaxons, that are characterised by well-defined relaxation times and, in the homogeneous case, also by proper velocities and mean free paths.
Thermal transport can thus be described as a kinetic theory of a gas of relaxons.
The characterisation of relaxon properties provides a description of the thermal transport in terms of proper time scales, and in the steady-state homogeneous case of velocity and length scales.
For clarity, we report in Table \ref{table1} a summary of relaxon characteristics and how they compare with phonons.
This theory is applied here first to graphene at room temperature where, as is typical of 2D materials or of 3D solids at low temperatures, the failure of the SMA and of its picture of phonons as heat carriers becomes dramatic; and to silicon at room temperature, where, although the SMA yields reasonable thermal conductivities, the theory brings new insight in the microscopic interpretation of heat flux and its typical velocities.
Finally, we have shown that the Matthiessen rule is violated in the exact BTE, with significant consequences for all systems in which the SMA does not hold.
As a final remark, the concept of relaxons has been applied in this work in the context of phonons; however, similar arguments will hold for the electron BTE or other semiclassical transport models.

\section*{Acknowledgements}

We gratefully acknowledge F. Mauri and G. Fugallo for useful discussions; 
the Swiss National Science foundation under project ID 200021\_143636 and National Centre of Competence in Research MARVEL, the Max Planck - EPFL Center for Molecular Nanoscience and Technology, and the Swiss National Supercomputing Center CSCS under project ID s580.

\section*{Methods}
\subsection*{First-principles simulations}
Density-functional theory calculations have been performed with the Quantum-ESPRESSO distribution \cite{qe}, using the local-density approximation and norm-conserving pseudopotentials from the PSLibrary \cite{pslib}; for graphene a plane-wave cutoff of 90 Ry and a Methfessel-Paxton smearing of 0.02 Ry have been used and for silicon a plane-wave cutoff of 100 Ry.
Graphene is simulated with a slab geometry, using an optimized lattice parameter $a=4.607$ Bohr and a cell height $c=3a$; for silicon we find an optimized lattice parameter of 10.18 Bohr.
The Brillouin zone is integrated with a Gamma-centered Monkhorst-Pack mesh of 24$\times$24$\times$1 points for graphene and 12$\times$12$\times$12 for silicon.
Second and third order force constants are computed on meshes of 16$\times$16$\times$1 and 4$\times$4$\times$1 points  respectively for graphene and 8$\times$8$\times$8 and 4$\times$4$\times$4 for silicon, and are later Fourier-interpolated on finer meshes.

\subsection*{Thermal conductivity simulations}
The scattering matrix $\tilde{\Omega}$ includes 3-phonon interactions and harmonic isotopic scattering \cite{fugalloPRB,marzari:prl} at natural abundances \cite{iupac} (98.93\% $^{12}$C, 1.07\% $^{14}$C for carbon, and 92.22\% $^{28}$Si, 4.67\% $^{29}$Si, 3.09\% $^{30}$Si  for silicon).
For graphene, the scattering matrix is constructed using the same computational parameters of Ref. \cite{fugallo-nano} (a Gaussian smearing of 10 cm$^{-1}$ and a mesh of 128$\times$128$\times$1 points for integrating the Brillouin zone), resulting in a matrix of order 98304, while for silicon we use a Gaussian smearing of 7 cm$^{-1}$ and a mesh of 30$\times$30$\times$30, yielding a matrix of order 162000.
$\tilde{\Omega}$ is diagonalised exactly using the routine PDSYEV of the Scalapack library \cite{scalapack}.
The simulation cell of graphene is renormalized using the interlayer distance of bulk graphite ($c/a=1.367$), in order to have a thermal conductivity comparable with the 3D counterpart.
We verified the correctness of the software implementation ensuring that the thermal conductivity estimated with the diagonalization solver coincides with that computed with the variational method of Ref. \cite{fugalloPRB} up to at least 4 significant digits.
It's worth mentioning that these calculations are not prohibitively expensive and could be extended to other systems. 
The present software implementation computes $\tilde{\Omega}$, diagonalizes it and computes the conductivity of graphene in about 5 hours using 256 CPUs on the Piz Daint supercomputer of the Swiss National Supercomputer Center (CSCS), for a total of 1300 CPU hours (1000 of which are spent in the diagonalization).
For silicon the calculation completed in 8 hours on 576 CPUs, for a total of 4600 CPU hours.
Calculations have been managed using the AiiDA materials' informatics platform \cite{AiiDA}

\section*{Appendix A: Scattering rates}
In this appendix we report the expressions for building the scattering matrix using 3-phonon and isotope scattering events, which are discussed in detail in Ref. \cite{fugalloPRB}.
To make a connection with other studies, we note that most contemporary ones have largely preferred to solve the BTE using a phonon deviation from equilibrium of the form $n_{\mu} = \bar{n}_{\mu} + \bar{n}_{\mu} (\bar{n}_{\mu}+1) F_{\mu}$.
Since the action of the collision operator must not change, we have the relation:
\begin{equation}
\sum_{\mu'} A_{\mu\mu'}F_{\mu'} = \sum_{\mu'} \Omega_{\mu\mu'} \Delta n_{\mu'} \;,
\end{equation}
where $A$ is the scattering matrix when it acts on $F$, related with the scattering matrices used in our work by
\begin{gather}
\Omega_{\mu\mu'}   = A_{\mu\mu'} \frac{1}{\bar{n}_{\mu'} (\bar{n}_{\mu'}+1)} \;, \label{omega_to_a} \\
\tilde{\Omega}_{\mu\mu'}  = \frac{1}{\sqrt{\bar{n}_{\mu} (\bar{n}_{\mu}+1)}}  A_{\mu\mu'} \frac{1}{\sqrt{\bar{n}_{\mu'} (\bar{n}_{\mu'}+1)}} \label{omegat_to_a} \;.
\end{gather}
The scattering rate for a phonon coalescence event is:
\begin{align}
P_{\mu\mu'}^{\mu''} =& \frac{2\pi}{N\hbar^2} \sum_{\boldsymbol{G}} |V^{(3)}(\mu,\mu',-\mu'')  |^2 \nonumber \\
& \times \bar{n}_{\mu} \bar{n}_{\mu'}(\bar{n}_{\mu''}+1) \delta_{\boldsymbol{q} + \boldsymbol{q}' - \boldsymbol{q}'', \boldsymbol{G}} \nonumber \\
& \times \delta(\hbar\omega_{\mu}+\hbar\omega_{\mu'}-\hbar\omega_{\mu''}) \;,
\end{align}
where $N$ is the number of $\boldsymbol{q}$ points, $\boldsymbol{G}$ is a reciprocal lattice vector and $V^{(3)}$ is the third derivative of the unit cell energy $\mathcal{E}^{\text{cell}}$ with respect to atomic displacements
\begin{equation}
V^{(3)}(\mu,\mu',\mu") = \frac{\partial^3 \mathcal{E}^{\text{cell}}} {\partial X_{\mu} \partial X_{\mu'} \partial X_{\mu''}} \;,
\end{equation}
with
\begin{equation}
X_{\boldsymbol{q}s} = \frac{1}{N} \sum_{l b \alpha} \sqrt{\frac{2M_b\omega_{\boldsymbol{q}s}}{\hbar}}
z_{\boldsymbol{q}s}^{*,b\alpha} u_{b\alpha}(\boldsymbol{R}_l) e^{-i\boldsymbol{q}\cdot \boldsymbol{R}_l} \;,
\end{equation}
where $b$ is an index running on the basis of atoms in the unit cell, $\boldsymbol{R}_l$ is a Bravais lattice vector identifying the $l^{\text{th}}$ unit cell inside the crystal, $\alpha$ is a cartesian index, $M_b$ is the mass of atom $b$, $z$ is the phonon polarization vector and $u$ is the vector of atomic displacements.
The scattering rate for a phonon-isotope scattering event is:
\begin{align}
P_{\mu\mu'}^{\text{isot}} =& \frac{\pi}{2N} \omega_{\mu} \omega_{\mu'} \bigg[  \bar{n}_{\mu}\bar{n}_{\mu'} + \frac{\bar{n}_{\mu}+\bar{n}_{\mu'}}{2} \bigg]  \nonumber \\
&\times \sum_b g_b \bigg| \sum_{\alpha} z_{\mu}^{*,b\alpha} z^{b\alpha}_{\mu'} \bigg|^2 \delta(\omega_{\mu} - \omega_{\mu'}) \;,
\end{align}
where $g_b = \frac{  \big(M_b - \big<M_b\big>\big)^2  }{   \big<M_b\big>^2 }$.

Combining these scattering rates, the scattering matrix $A$ is:
\begin{align}
A_{\mu \mu'} =&  \bigg[   \sum_{\mu'',\mu'''} \bigg( P_{\mu\mu'''}^{\mu''}+\frac{1}{2}P_{\mu''\mu'''}^{\mu} \bigg)  + \sum_{\mu'} P_{\mu\mu'}^{\text{isot}} \bigg] \delta_{\mu\mu'} \nonumber \\
&- \sum_{\mu''} (P_{\mu\mu''}^{\mu'} - P_{\mu\mu'}^{\mu''} + P_{\mu'\mu''}^{\mu}) + P_{\mu\mu'}^{\text{isot}} \;.
\end{align}
In the numerical implementation, the Dirac delta conserving the energy is replaced by a Gaussian smearing.
As the authors of Ref. \cite{fugalloPRB} noted, the above expression guarantees that the scattering matrix $A$ is symmetric and positive-definite also in presence of a Gaussian smearing (other expressions, which would be equivalent with a Dirac delta function, may introduce spurious negative eigenvalues).
By virtue of Eqs. (\ref{omega_to_a}) and (\ref{omegat_to_a}), it follows that $\tilde{\Omega}$ is symmetric but not $\Omega$, hence the necessity of the transformations (\ref{n_symmetrisation}) and (\ref{n2_symm}).

Finally, we recall that phonon lifetimes are related to the diagonal elements of the scattering matrix as:
\begin{gather}
A_{\mu\mu} = \frac{\bar{n}_{\mu} (\bar{n}_{\mu}+1) }{\tau_{\mu}} \;, \\
\tilde{\Omega}_{\mu\mu} = \frac{1 }{\tau_{\mu}} \;.
\end{gather}

\section*{Appendix B: silicon thermal properties}

In this appendix we study the effect of the scaling of specific heat on velocities and mean free paths.
While relaxon properties are as defined in the main text, phonon velocities and mean free paths are scaled as:
\begin{gather}
v_{\mu} \to \sqrt{\frac{C_{\mu}}{\mathcal{V}C}} v_{\mu}  \;,\\
\Lambda_{\mu}^{\text{SMA}} \to \sqrt{\frac{C_{\mu}}{\mathcal{V}C}} \Lambda_{\mu}^{\text{SMA}} \;.
\end{gather}
With this choice of scaling, we can write the SMA thermal conductivity as:
\begin{equation}
(k^{ij})^{\text{SMA}} = \sum_{\mu} C v_{\mu}^i (\Lambda_{\mu}^j)^\text{SMA} \; ,
\end{equation}
which treats specific heat in the same way as Eq. \ref{k_relaxon}.
\begin{figure}
\centering
\includegraphics{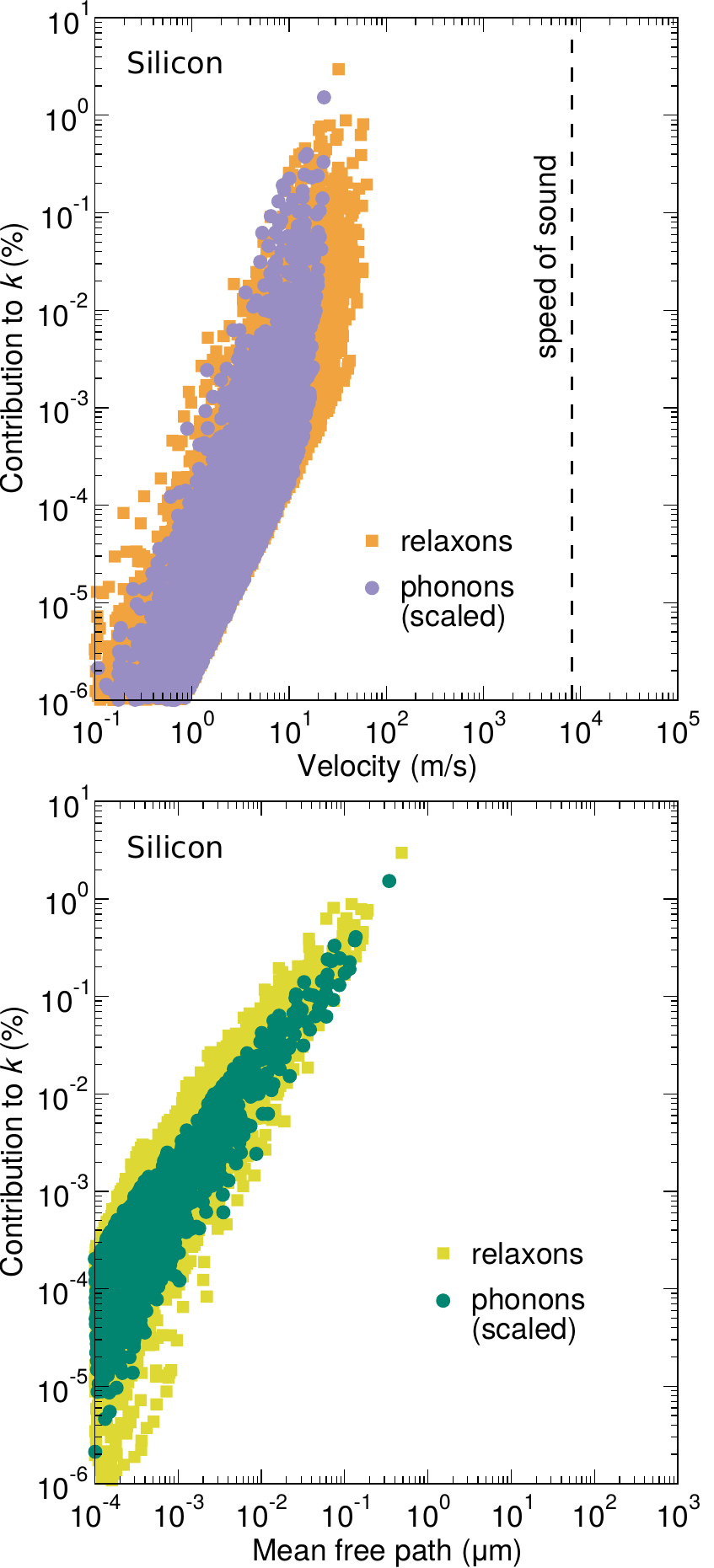}
\caption{
Top panel: comparison of relaxon velocities with scaled phonon velocities. Bottom panel: same for mean free paths.
}
\vspace{0.1cm}
\label{figsi_scaled}
\end{figure}
In Figure \ref{figsi_scaled} we report the comparison of these scaled phonon quantities with the relaxon properties in silicon.
One can readily see that the 2 orders of magnitude of difference that appeared in Figs. \ref{fig4_si} and \ref{fig5_si} have almost disappeared.
Most of the discrepancy is thus due to the usage of specific heat.
Nevertheless, the largest phonon velocities are still a factor of 3 smaller than those of relaxons, and the two pictures do not perfectly overlap.

\section*{Appendix C: Matthiessen rule}

Here we recall a known result [18] that proves that the application of the Matthiessen rule results in an overestimation of the exact thermal conductivity.
The BTE for a homogeneous system under a static gradient of temperature can be written in a matrix form (see for example Ref. [18]\ or more recently Ref. [7]):
\begin{equation}
A \phi = b \;, 
\end{equation}
where $A$ is related to the scattering matrix $\Omega$ via $A_{\mu\mu'} = \Omega_{\mu\mu'} \bar{n}_{\mu'}(\bar{n}_{\mu'}+1)$, $b=-\frac{\partial \bar{n}_{\mu}}{\partial T} v_{\mu}$ and $\phi$ is the deviation from equilibrium defined as $n_{\mu}=\bar{n}_{\mu}+\bar{n}_{\mu}(\bar{n}_{\mu}+1)\nabla T \phi_{\mu}$.

Another way of solving the BTE, besides the diagonalization approach discussed in the main article, is via the variational principle [18].
In particular, the solution of the BTE can be found from the minimization of the functional [18]
\begin{equation}
\mathcal{F}[\phi] = \frac{ \big< \phi \big|  A \big| \phi \big> }{ (\big< \phi \big|  b \big>)^2 } \;.
\end{equation}
Let $\phi$ be the function minimising $\mathcal{F}$.
The minimum of  $\mathcal{F}$ is directly proportional to the thermal resistivity $\rho$ [18]; therefore we write
\begin{equation}
\rho = \frac{1}{k} = \mathcal{F}[\phi] \;.
\end{equation}
Now, let's separate the scattering matrix into two different components (for example  3-phonon and isotopic scattering)
\begin{equation}
A = A_1 + A_2 \;.
\end{equation}
The exact resistivity is given by:
\begin{equation}
\rho = \frac{ \big< \phi \big|  A_1 \big| \phi \big> + \big< \phi \big|  A_2 \big| \phi \big> }{ (\big< \phi \big|  b \big>)^2 } \;.
\end{equation}
The function $\phi$ that minimises the functional defined by $A$ will not be, in general, the function that minimises the functionals $\mathcal{F}_1$ and $\mathcal{F}_2$ defined by $A_1$ or $A_2$ only.
The functionals $\mathcal{F}_1$ and $\mathcal{F}_2$ instead will be minimised by the functions $\phi_1$ and $\phi_2$ respectively.
By the variational principle
\begin{align}
\rho 
&= \frac{ \big< \phi \big|  A_1 \big| \phi \big> }{ (\big< \phi \big|  b \big>)^2 } 
+  \frac{ \big< \phi \big|  A_2 \big| \phi \big> }{ (\big< \phi \big|  b \big>)^2 } \\
&\ge \frac{ \big< \phi_1 \big|  A_1 \big| \phi_1 \big> }{ (\big< \phi_1 \big|  b \big>)^2 } 
+ \frac{ \big< \phi_2 \big|  A_2 \big| \phi_2 \big> }{ (\big< \phi_2 \big|  b \big>)^2 }  \nonumber \\
&= \rho_1 + \rho_2 \nonumber \;.
\end{align}
Alternatively, this can be written as
\begin{equation}
\frac{1}{k} \le \frac{1}{k_1} + \frac{1}{k_2} \;,
\end{equation}
showing that the Matthiessen rule is a special case where the equalities holds exactly. More generally, its application leads to an overestimation of thermal conductivities.

\section*{Appendix D: iterative method}

\begin{figure}
\centering
\includegraphics{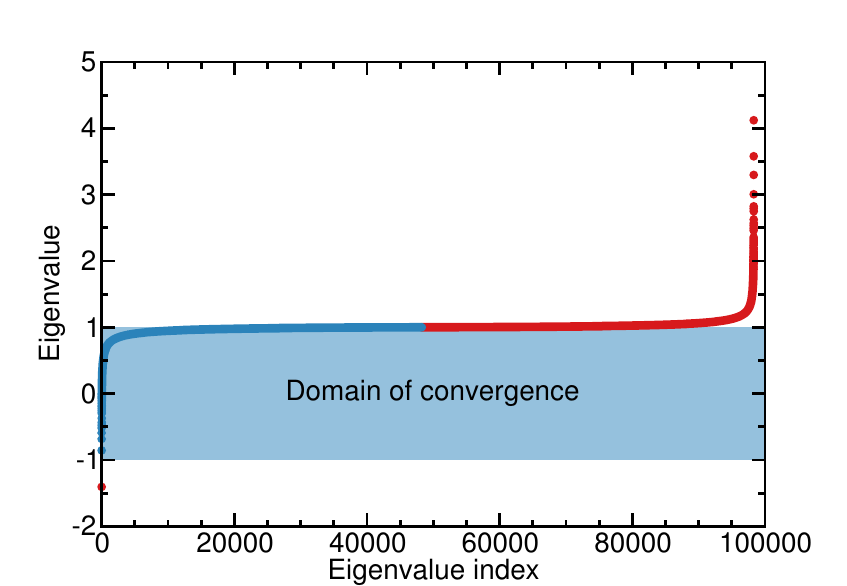}
\caption{
Eigenvalues $\lambda$ of the matrix $A_{\text{d}}^{-1} A_{\text{od}}$ (see main text for the definition), for graphene at room temperature, ordered by their magnitude. The red dots, roughly half of the eigenvalue spectrum, indicate eigenvalues $|\lambda|>1$ that cause a divergence of the iterative solution for the Boltzmann transport equation. Most of the unstable eigenvalues are greater than 1, with only one eigenvalue lower than -1.
}
\vspace{0.1cm}
\label{fig7}
\end{figure}

In this appendix we examine the convergence properties of the iterative method for solving the BTE.
Such method can be formalised as follows.
The steady state homogeneous BTE in the phonon basis is:
\begin{equation}
\nabla T \cdot {\bf v}_{\mu} \frac{\partial n_{\mu}}{\partial T}  = - \frac{1}{\mathcal{V}} \sum_{\mu'} \Omega_{\mu\mu'} n_{\mu'} \; .
\end{equation}

This is a linear algebra problem of the form $AF=b$, where $b_{\mu}=-v_{\mu} \hbar \omega_{\mu} \bar{n}_{\mu}(\bar{n}_{\mu}+1)$, $F$ is defined by $n_{\mu}=\bar{n}_{\mu} + \bar{n}_{\mu}(\bar{n}_{\mu}+1)\nabla T F_{\mu}$ and $\Omega_{\mu\mu'} = A_{\mu\mu'} \bar{n}_{\mu'}(\bar{n}_{\mu'}+1)$.
The iterative solution for $F$ [4,16,17] can then be recast [7] as a geometric series $F = \sum_{j=0}^{\infty} \big[ - (A^d)^{-1} A^{od} \big]^j  (A^d)^{-1} b$, where $A^{d}$ and $A^{od}$ are respectively the diagonal and the off-diagonal parts of $A$.
This series is convergent if and only if all the eigenvalues $\lambda$ of $\big[(A^d)^{-1} A^{od}\big]$ are $|\lambda|<1$.
In Fig. \ref{fig7}, we show that in graphene $|\lambda|>1$ for more than half of the spectrum, proving that the iterative method is numerically unstable for graphene at room temperature.
In general, one might expect convergence issues for the iterative method whenever the relaxon picture differs significantly from the phonon picture and the contribution of the off-diagonal part is large compared to the diagonal part.


%

\end{document}